\documentclass[runningheads]{svmult}

\usepackage{makeidx}   
\usepackage{graphicx}  
\usepackage{subeqnar}  
\usepackage{multicol}  
\usepackage{physprbb}  
\makeindex             

\newcommand{\epm}{\ensuremath{e^{\pm}\;}}

\def\eg{{\it e.g.},~}
\def\etal{{\it et al.}~}
\def\4he{$^4$He}
\def\3he{$^3$He}
\def\7li{$^7$Li}
\def\Yp{Y$_{\rm P}$~}
\def\yd{$y_{\rm D}$~}

\def\hii{H\thinspace{$\scriptstyle{\rm II}$}~}

\def\popii{Pop\thinspace{$\scriptstyle{\rm II}$}~}

\newcommand\la{\lower0.6ex\vbox{\hbox{\ensuremath{\buildrel{\textstyle<}\over{\sim}\ }}}}
\newcommand\ga{\lower0.6ex\vbox{\hbox{\ensuremath{\buildrel{\textstyle>}\over{\sim}\ }}}}

\begin{document}
\title*{BBN and the Primordial Abundances}
%
%
%
%
\titlerunning{BBN and the Primordial Abundances}
%
\author{Gary Steigman}
\authorrunning{Gary Steigman}
%
%
\institute{The Ohio State University, Columbus OH 43210, USA}

\maketitle              

\begin{abstract}
The relic abundances of the light elements synthesized during the first 
few minutes of the evolution of the Universe provide unique probes of 
cosmology and the building blocks for stellar and galactic chemical 
evolution, while also enabling constraints on the baryon (nucleon) 
density and on models of particle physics beyond the standard model.  
Recent WMAP analyses of the CBR temperature fluctuation spectrum, 
combined with other, relevant, observational data, has yielded very 
tight constraints on the baryon density, permitting a detailed, 
quantitative confrontation of the predictions of Big Bang Nucleosynthesis
with the post-BBN abundances inferred from observational data.
The current status of this comparison is presented, with an emphasis 
on the challenges to astronomy, astrophysics, particle physics, and
cosmology it identifies.

\end{abstract}

\section{Introduction and Overview}
Our Universe is observed to be expanding and filled with
radiation (the Cosmic Background Radiation: CBR), along with ``ordinary"
matter (baryons $\equiv$~nucleons).  It is well known that if this evolution
is followed backwards in time, then there was an epoch during its early
evolution when the Universe was a ``Primordial Nuclear Reactor", synthesizing
in astrophysically interesting abundances the light nuclides D, \3he, \4he,
and \7li.  Discussion of BBN can start when the Universe was some tens of
milliseconds old and the temperature (thermal energies) was of order a 
few MeV.  At that time there were no complex nuclei, only neutrons and
protons.  Since there are nearly ten orders of magnitude more CBR photons
in the Universe than nucleons, photodissociation ensures that at such 
high temperatures the abundances of complex nuclei are negligibly small.  
However, as the Universe expands (and the weak interactions transmute 
neutrons and protons), collisions among nucleons begin to build the light 
nuclides when the temperature drops below $\sim 100$~MeV, and the Universe 
is a couple of minutes old.  Very quickly, almost all neutrons available 
are incorporated in the most tightly bound of the light nuclides, \4he.  
As a result, the \4he primordial abundance (mass fraction: Y$_{\rm P}$) 
is a sensitive probe of the competition between the weak interaction rates 
and the universal expansion rate (the Hubble parameter: $H$); \4he is 
a cosmological chronometer.  The reactions building \4he are not rate 
(nuclear reaction rate) limited and, therefore, \Yp is only weakly 
(logarithmically) sensitive to the baryon density.  In contrast, the 
BBN abundances of the other light nuclides (D, \3he, \7li) are rate 
limited and these do depend sensitively (to lesser or greater degrees) 
on the baryon density; D, \3he, and \7li are all potential baryometers.

The relic abundances of the light nuclides predicted by BBN in the 
``standard" model of cosmology (SBBN) depend only on one free parameter: 
the nucleon density.  There are, therefore, two complementary approaches 
to testing SBBN.  On the one hand, the primordial abundances inferred
from observational data should be consistent with the SBBN predictions
{\bf for a unique value (or range) of the nucleon density}.  On the 
other hand, if there is a non-BBN constraint on the range allowed for 
the baryon density, this should lead to SBBN-predicted abundances
in agreement with the observational data.  Is this the case?  If not,
why not?  That is, if there are conflicts between predictions and
observations, is the ``blame" to be laid at the feet of the observers
(inaccurate data and/or unidentified systematic errors?), or the
astrophysicists (incorrect models for analysing the data and/or
extrapolating from abundances determined at present (``here and now")
to their primordial (``there and then") values), or are the standard
models of particle physics and/or cosmology in need of revision?

\subsection{The Status Quo: Observations Confront SBBN}
\begin{figure}[b]
\begin{center}
\includegraphics[width=.5385\textwidth]{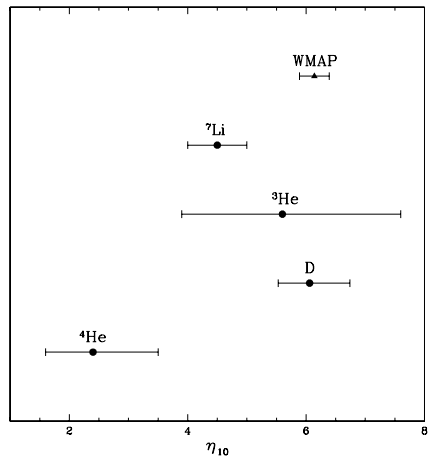}
\end{center}
\caption[]{The baryon density parameter, $\eta_{10}$, inferred from 
SBBN and the relic abundances of D, \3he, \4he, and \7li (filled 
circles), along with the non-BBN determination from WMAP (filled 
triangle).  See the text for details.}
\label{eta}
\end{figure}

Space limitations prevent my presenting a full-fledged review of 
the history of the observational programs along with the evolution 
of the comparisons between theory and data.  For some recent reviews
of mine, see~\cite{steig1} and further references therein.  Instead,
an overview is presented which highlights the challenges to SBBN.  
The remainder of this article is devoted to some of the key issues 
associated with each of the light nuclide relic abundances.

If the standard model (SBBN) is the correct choice and if the primordial 
abundances inferred from the data were free of systematic errors, then
the baryon densities determined from D, \3he, \4he, and \7li should agree
among themselves and with that inferred from the CBR (WMAP) and other,
non-BBN, cosmological data.  In Figure 1 are plotted the various values
of the baryon density parameter\footnote{After \epm annihilation during 
the early evolution of the Universe, the ratio of baryons to photons 
is, to a very good approximation, preserved down to the present.  The 
baryon density parameter is defined to be this ratio (at present):
$\eta \equiv n_{\rm N}/n_{\gamma}$; $\eta_{10} \equiv 10^{10}\eta$.}
determined by SBBN and the adopted primordial abundances and, also,
from the WMAP-team analysis~\cite{sperg}.  As may be seen from Fig.~1,
the SBBN D abundance is in excellent agreement with the WMAP-inferred 
baryon density.  However, neither \7li nor \4he agree with them or, 
even with each other.  While \3he is consistent with SBBN deuterium 
and with WMAP (and is not in disagreement with \7li), the very large 
uncertainty in its primordial abundance, combined with its relative 
insensitivity to the baryon density, render it -- at present -- an 
even less sensitive baryometer than is \4he.

In the next section each light nuclide is considered in turn, its 
post-BBN evolution briefly reviewed along with identification of a 
few of the potential challenges to accurately inferring the primordial 
abundances from the observational data.  Then, having established that 
the current data -- taken at face value -- are not entirely consistent 
with SBBN, I investigate whether changes in the early universe expansion 
rate can reconcile them.

\section{Primordial Abundances -- Evolution, Uncertainties, Systematics}

While Figure 1 suggests some problems with SBBN, the optimist may 
prefer to conclude that observations have provided impressive 
confirmation of the standard cosmological model.  After all, if
the standard model -- or something very much like it -- were not
correct, there'd be no good reason why the abundances of four 
light nuclides, which range over some nine orders of magnitude, 
should be just such that the baryon densities inferred from each 
of them lie within a factor of three of each other, in nearly 
perfect agreement with that derived independently from non-BBN 
data.  Only recently, when cosmology has entered an era of great
precision, has it become important to distinguish accuracy from
precision and to revisit the path from precise astronomical observations
to accurate abundances.  For each of the light nuclides of interest
here, an all too abbreviated overview of the current uncertainties
is presented, with the intentional goal of creating controversey 
in order to stimulate future observations and theoretical analyses.

\subsection{Deuterium}

\begin{figure}[b]
\begin{center}
\includegraphics[width=.5385\textwidth]{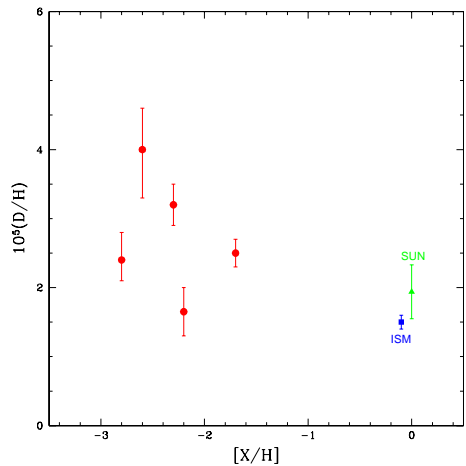}
\end{center}
\caption[]{The deuterium abundance (by number relative to hydrogen), 
$y_{\rm D} \equiv 10^{5}$(D/H), derived from high redshift, low
metallicity QSOALS~\cite{kirk}~(filled circles).  The metallicity 
is on a log scale relative to solar; depending on the line-of-sight, 
$X$ may be oxygen or silicon.  Also shown is the solar system 
abundance (filled triangle) and that from observations of the 
local ISM (filled square).}
\label{d}
\end{figure}

Deuterium is the baryometer of choice.  During its post-BBN evolution,
as gas is cycled through stars, D is only destroyed (setting aside
rare astronomical events where, far from equilibrium, tiny amounts
of D may be synthesized).  Therefore, observations of D anywhere,
at any time (\eg the solar system or the local ISM), provide a 
{\it lower} bound to its primordial abundance.  As a result, it is
expected that observations of D in regions which have experienced
minimal stellar evolution (\eg the high redshift and/or low metallicity
QSO Absorption Line Systems: QSOALS) should provide a good estimate 
of the relic abundance of deuterium.  Kirkman \etal \cite{kirk} have
gathered the extant data; see \cite{kirk} for details and related
references.  In Figure 2 are shown the QSOALS deuterium abundances
as a function of metallicity; for reference, solar system and ISM
D abundances are also shown.

While the observers are to be commended for their heroic work in
identifying and analysing the tiny fraction of QSOALS which can be 
used to infer a low-metallicity, high redshift, D abundance, there 
are several unsettling aspects of the data displayed in Figure 1. 
Perhaps most noticeable is the paucity of data points.  Without at 
all minimizing the difficulty of finding and analysing such systems,
it is very nearly a sin to claim that the primordial abundance of
a cosmologically key light nuclide is determined by five data points.  
If, however, the data points were in agreement within the estimated 
statistical errors, this might be less disturbing.  It is clear from 
Fig.~1 that this is not the case.  For these five data points the 
$\chi^{2}$ about the mean is $\ga 16$, suggesting either that the 
uncertainties have been underestimated, or that some of these data 
may be contaminated by unidentified systematic errors.  While the 
dispersion may simply be due to the small number of data points, 
it might be significant that the three QSOALS with the lower D/H 
ratios are LLS, while the two highest D/H determinations are from 
DLAs.  

In the absence of evidence for changing or eliminating any of the 
current D abundance determinations, it is not unreasonable to follow 
the advice of \cite{kirk} and adopt the weighted mean as an estimate 
of the primordial D abundance.  Here, too, there is a (minor) problem.  
Kirkman \etal \cite{kirk} advocate finding the mean of log(D/H) to 
determine $y_{\rm D} \equiv 10^{5}($D/H); \yd $\equiv 10^{(5+<{\rm 
log(D/H)}>)} = 2.78$.  In contrast, if the weighted mean of the five 
D/H determinations is used, \yd = 2.60.  While this difference is 
well within the dispersion, it reflects the fact that in determining 
the mean of log(D/H), PKS1937 with \yd = 3.24 dominates, while for 
the mean of D/H, HS0105 with \yd = 2.54 dominates.  In what follows 
I adopt \yd = 2.6$\pm0.4$, where the error, following \cite{kirk}, is 
derived from the dispersion in D/H determinations.  The corresponding 
BBN (SBBN) prediction for the baryon abundance, $\eta_{10}({\rm D}) 
\approx 6.1^{+0.7}_{-0.5}$, is shown in Figure 1.

\begin{figure}[b]
\begin{center}
\includegraphics[width=.5385\textwidth]{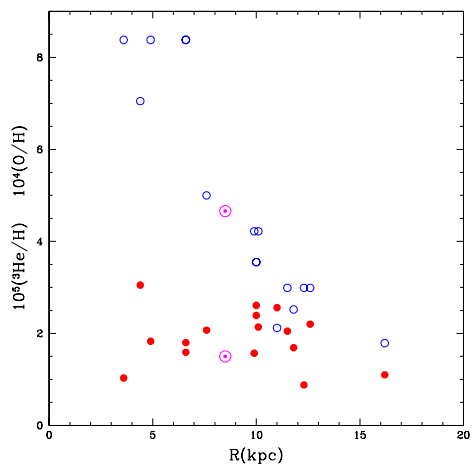}
\end{center}
\caption[]{The \3he abundances (by number relative to hydrogen), 
$y_{3} \equiv 10^{5}(^{3}$He/H), derived from Galactic \hii 
regions~\cite{rood}, as a function of galactocentric distance 
(filled circles).  Also shown for comparison is the solar system 
abundance (solar symbol).  The open circles are the oxygen 
abundances for the same \hii regions (and for the Sun).}
\label{3he}
\end{figure}

\subsection{Helium-3}

In my talk at this meeting I actually avoided discussion of \3he,
until it was forced upon me during the question session.  In part,
this was because this subject was ably covered in Tom Bania's talk,
in Dana Balser's poster, and in Bob Rood's conference summary.  In
part, however, it was because, in my opinion, for both observational 
and theoretical reasons \3he has more to teach us about stellar and
Galactic evolution than about BBN.  \3he is a less sensitive baryometer
than is D since (D/H)$_{\rm BBN} \propto \eta^{-1.6}$, while 
$(^{3}$He/H)$_{\rm BBN} \propto \eta^{-0.6}$.  Even more important is 
that, in contrast to the expected, monotonic, post-BBN evolution of D, 
the post-BBN evolution of \3he is quite complicated, with competition 
between stellar production, destruction, and survival.  For years,
indeed decades, it had been anticipated that net stellar production
would ``win" and the abundance of \3he would increase with time (and
with metallicity); see, \eg \cite{rst}.  Unfortunately, observations 
of \3he are restricted to the solar system and the Galaxy.  Nonetheless,
since there is a clear galactic gradient of metallicity (see Fig.~\ref{3he}), 
a gradient in \3he abundance would also be expected.  If net production 
``wins", then \3he should be highest in the inner galaxy; the opposite 
if net destruction dominates.  The absence of any statistically 
significant gradient in the Bania, Rood, Balser (BRB) data~\cite{rood} 
(see Fig.~\ref{3he}), points to an extremely delicate balance (cancellation) 
between production and destruction.  This suggests that the mean 
\3he abundance in the Galaxy ($y_{3} \approx 1.9$) might provide 
a reasonable estimate of the primordial abundance.  However, BRB 
recommend that the \3he abundance determined in the most distant 
(from the Galactic center), most metal poor Galactic \hii region 
yields an upper limit, $y_{3}~\la 1.1\pm0.2$, to the primordial 
abundance.  The estimate of $\eta_{10}({^{3}{\rm He}}) \approx 
5.6^{+2.0}_{-1.7}$ shown in Figure 1 is based on this choice.  
Had I adopted the mean value of $y_{3} = 1.9$, I would have 
inferred $\eta_{10}(^{3}{\rm He}) \approx 2.3$.  While the former 
choice is in excellent agreement with deuterium (and with \7li 
and with the WMAP result~\cite{sperg}), the very large uncertainty 
renders \3he an insensitive baryometer; the latter option would 
be consistent with \4he, but not with D (or with \7li or WMAP).

\subsection{Helium-4}

\begin{figure}[b]
\begin{center}
\includegraphics[width=.5385\textwidth]{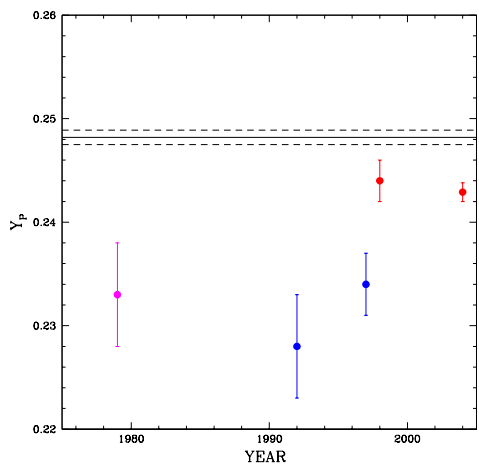}
\end{center}
\caption[]{A summary of the time evolution of primordial \4he abundance 
determinations (mass fraction Y$_{\rm P}$) from observations of 
metal-poor, extragalactic \hii regions (see the text for references).  
The solid horizontal line is the SBBN-predicted \4he abundance 
expected for the WMAP (and/or D) inferred baryon density.  The 
two dashed lines show the $1\sigma$ uncertainty in this prediction.}
\label{4he}
\end{figure}

Helium-4 is the textbook example of a relic nuclide whose abundance is
known precisely but, likely, inaccurately.  To be of value in testing
SBBN as well as in constraining non-standard models, \Yp should be determined
to 0.001 or better.  The largest uncertainty in the SBBN prediction is
from the very small error in the neutron lifetime ($\tau_n = 885.7\pm0.7$~s).
For the WMAP estimate of the baryon density, including its uncertainty,
the SBBN-predicted primordial abundance is \Yp $= 0.2482\pm0.0007$, as
shown in Figure~\ref{4he}.  Also shown in Figure~\ref{4he} is a record of 
\Yp determinations, from observations of extragalactic, low metallicity,
\hii regions, covering the period from the late 1970s to the present (2004).
During this time it has been well known, but often ignored, that there
are a variety of systematic uncertainties which might dominate the \Yp
determinations.  In the hope of accounting for these systematic errors
(rather than constraining them by observations), the error estimates for
\Yp have often been inflated beyond the purely statsitical uncertainties.
Thus, until the late 1990s, the typical error estimate for \Yp was 0.005.
For example, summarizing the status as of 2000, Olive, Steigman, and 
Walker~\cite{osw} suggested that the data available at that time were 
consistent with \Yp $= 0.238\pm0.005$; see Fig.~\ref{4he}.  However, if 
\4he were used as an SBBN baryometer (not recommended!), the error in 
the baryon density would have been $\sim 50$\%.  Of course, as the number 
of \hii regions observed increased, largely due to the work of Izotov \& 
Thuan~\cite{hevstime}, the statistical errors decreased.  For example, 
from observations of 82 extragalactic \hii regions, in their 2004 paper 
Izotov \& Thuan quote~\cite{hevstime} \Yp $= 0.2429\pm0.0009$; this data 
point is shown in Fig.~\ref{4he}.  In contrast, a very recent, detailed 
study of the effects of {\it some} of the identified systematic uncertainties 
by Olive \& Skillman~\cite{os} suggests the true errors are likely larger 
than this, by at least an order of magnitude.  

As may be seen from Fig.~\ref{4he}, none of the \Yp estimates agree with 
the SBBN prediction, all being low by roughly 2-$\sigma$.  Indeed, from 
their sample of 82 data points Izotov \& Thuan~\cite{hevstime} derive 
such a small uncertainty that their central value is low by nearly
6-$\sigma$.  In their analysis, Izotov \& Thuan commit the cardinal sin of
examining their data and then, {\it a posteriori}, choosing a subsample of 
7 \hii regions to derive their favored estimate of \Yp $= 0.2421\pm0.0021$.  
One wonders what they may have found from a random series of choosing 7 of 
82 data points.  In any case, this estimate also falls short of the SBBN 
prediction by nearly 3-$\sigma$.  Using this suspect subsample, Olive \& 
Skillman~\cite{os} do find consistency with SBBN once they have corrected 
for the systematic errors they've chosen to include.  However, one they have 
ignored, the ionization correction factor~\cite{icf}, almost certainly would 
have the effect of reducing their central value and increasing their error 
estimate.  For the entire Izotov \& Thuan sample, Olive \& Skillman find 
a very large range for Y$_{\rm P}$, from 0.232 to 0.258 (or, \Yp $\approx 
0.245\pm0.013$, entirely consistent with SBBN)).  If the average correction 
for ionization suggested by Gruenwald, Steigman, and Viegas~\cite{icf} for 
the somewhat smaller 1998 Izotov \& Thuan data set is applied to their 2004 
compilation, it would suggest the Olive \& Skillman central value be reduced 
and their error increased: \Yp $\approx 0.239\pm0.015$.  

If, indeed, the true uncertainty in \Yp is really this large, it opens the 
possibility that there might be alternate approaches to \Yp which are 
competitive with the extragalactic, low-metallicity \hii regions and, even 
more important, complementary in that they are subject to completely 
independent sets of systematic errors.  One such example, with a venerable
history of its own, is to use effect of the initial helium abundance on the
evolution of low-mass \popii stars, employing the R-parameter~\cite{iben}.
Recently, Cassisi, Salaris, and Irwin~\cite{csi} have attempted this using
observations of a large sample of Galactic Globular Clusters (GGC) and new 
stellar models.  While they claim an extraordinarily accurate determination 
of \Yp ($0.243\pm0.006$), this does not seem to be supported by the data 
they present since, for the lowest metallicity GGCs, Y ranges from $\la 
0.19$ to $\ga 0.27$.  Nonetheless, if there is the possibility that the
R-parameter method might achieve theoretical and observational uncertainties
$\la 0.01$, it is certainly an approach worth pursuing.

An alternate approach, subject to large theoretical uncertainties, would be
to attempt to use chemical evolution models, tied to the solar helium and 
heavy element abundances, to extrapolate back to the relic abundance of \4he.
While at present this approach appears to be limited by the theoretical
uncertainties (\eg metallicity dependent stellar winds and stellar yields),
the following example may serve as a stimulus (or challenge) to those who 
might believe they can do better.  In a recent paper employing new yields, 
Chiappini, Matteucci, and Meynet~\cite{cmm} find $\Delta$Y$ \equiv $Y$_{\odot} 
- $Y$_{\rm P} \approx 0.018\pm0.006$.  Using the recent Bahcall \& 
Pinsonneault~\cite{bp04+} estimate of Y$_{\odot} \approx 0.270\pm0.004$, 
this leads to a primordial estimate of \Yp $\approx 0.252\pm 0.007$.  Although 
this result is consistent with the SBBN prediction, it would be entirely 
premature to declare victory on the basis of such a crude estimate.  The
possible lesson illustrated by this example is that the {\bf error} 
associated with such an approach might not be uncompetitive with those 
from the standard \hii region analyses.

\subsection{Lithium-7}

As with \4he, the recent history of the comparison between the SBBN
predictions and the observational data leading to the relic abundance
of \7li is one of conflict, with the spectre of systematic errors
looming large.  \7li, along with $^{6}$Li, $^{9}$Be, $^{10}$B, and
$^{11}$B, is produced in the Galaxy by cosmic ray spallation/fusion
reactions.  Furthermore, observations of super-lithium rich red
giants provide evidence that (some) stars are net producers of 
lithium.  Therefore, to probe the BBN yield of \7li, it is necessary
to restrict attention to the most metal-poor halo stars (the 
``Spite plateau").  Using a specially selected data set of the 
lowest metallicity halo stars, Ryan \etal \cite{ryan} claim evidence
for a 0.3 dex increase in the lithium abundance ([Li] $\equiv 12
+ $log(Li/H)) for $-3.5 \leq~$[Fe/H]$~\leq -1$, and they derive a
primordial abundance of [Li]$_{\rm P} \approx 2.0-2.1$.  This value is 
low compared to the estimate of Thorburn (1994)~\cite{thor}, who found
[Li]$_{\rm P} \approx 2.25\pm0.10$.  In the steps from the observed
equivalent widths to the derived abundances, the stellar temperature
plays a key role.  When using the infrared flux method effective 
temperatures, studies of halo and Galactic Globular Cluster stars
\cite{bonif} suggest a higher abundance: [Li]$_{\rm P} = 2.24\pm0.01$.
Very recently, Melendez \& Ramirez~\cite{mr} have reanalyzed 62 halo
dwarfs using an improved infrared flux method effective temperature
scale.  They fail to find the [Li] vs. [Fe/H] gradient claimed by
Ryan \etal \cite{ryan} and they confirm the higher lithium abundance,
finding  [Li]$_{\rm P} = 2.37\pm0.05$.  As shown in Figure 1, if
this were the true primordial abundance of \7li, then $\eta_{10}({\rm Li}) 
= 4.5\pm0.4$.  {\bf All} of these observational determinations of 
primordial lithium are significantly lower than the SBBN expectation 
of [Li]$_{\rm P} = 2.65^{+0.05}_{-0.06}$ for the WMAP baryon density.

As with \4he, the culprit may be the astrophysics rather than the
cosmology.  Since the low metallicity, dwarf, halo stars used to constrain
primordial lithium are the oldest in the Galaxy, they have had the most
time to modify (by dilution and/or destruction) their surface abundances.
While mixing of the stellar surface material with the interior would
destroy or dilute the prestellar lithium abundance, the very small
dispersion in [Li] among the low metallicity halo stars suggests this
effect may not be large enough to bridge the $\approx 0.3$~dex gap 
between the observed and WMAP/SBBN-predicted abundances ([Li]$^{obs}_{\rm P} 
\approx 2.37$ versus [Li]$^{pred}_{\rm P} \approx 2.65$); see, 
\eg \cite{pinsono} and further references therein.

\section{Non-Standard BBN}

\begin{figure}[b]
\begin{center}
\includegraphics[width=.5385\textwidth]{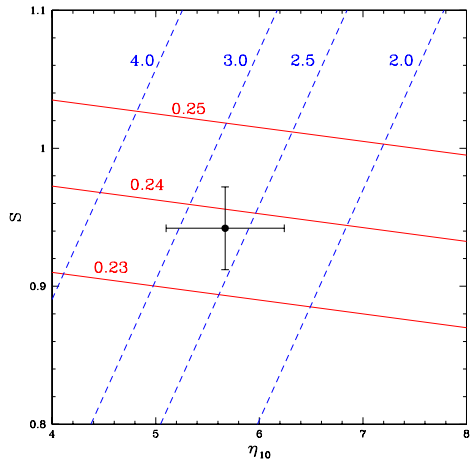}
\end{center}
\caption[]{Isoabundance curves for \4he (solid) and D (dashed) 
in the baryon abundance ($\eta_{10}$) -- expansion rate factor 
($S$) plane.  The labels on the \4he curves are for Y$_{\rm P}$, 
while those on the D curves are for $y_{\rm D} \equiv 10^{5}$(D/H).  
The filled circle with error bars corresponds to the adopted values 
of the D and \4he primordial abundances (see the text).}
\label{svseta}
\end{figure}

The path from acquiring observational data to deriving primordial
abundances is long and complex and littered with pitfalls.  While
the predicted and observed relic abundances are in rough qualitative 
agreement, at present there exist some potential discrepancies.  
These may well be due to the data, the data analysis, or the 
extrapolations from here and now to there and then.  But, there 
is also the possibility that these challenges to SBBN are providing 
hints of new physics beyond the standard models of particle physics 
and/or cosmology.  Once this option is entertained, the possibilities 
are limited only by the imagination and creativity of physicists and 
cosmologists.  Many such models have already been proposed and studied.  
One option is to discard most or all of standard physics and start 
afresh.  A more conservative (not in the pejorative sense!) approach 
is to recognize that the standard model (SBBN) does quite well and 
to look for small variations on the same theme.  Here, I'll adopt 
the latter strategy and explore one such option which can resolve 
some, but not all, of the conflicts: a non-standard, early universe 
expansion rate. 

There are many different extensions of the standard model of particle
physics which result in modifications of the early universe expansion 
rate (the time -- temperature relation).  For example, additional particles 
will increase the energy density (at fixed temperature), resulting in 
a faster expansion.  In such situations it is convenient to relate the 
extra energy density to that which would have been contributed by an 
additional neutrino with the ordinary weak interactions~\cite{ssg}.  
Just prior to \epm annihilation, this may be written as
\begin{equation}
{\rho' \over \rho} \equiv 1 + {7\Delta{\rm N}_{\nu} \over 43}.
\end{equation} 
Since the expansion rate (the Hubble parameter) depends on the square
root of the combination of G (Newton's contstant) and the density, 
the expansion rate factor ($S$) is related to $\Delta$N$_{\nu}$ by,
\begin{equation}
S \equiv {H' \over H} = (1 + {7\Delta{\rm N}_{\nu} \over 43})^{1/2} 
= ({G' \over G})^{1/2}.
\end{equation}
Thus, while adding new particles (increasing the energy density) 
results in a speed-up in the expansion rate ($\Delta$N$_{\nu} > 0$;
$S > 1$), changing the effective Newton's constant may either increase
or decrease the expansion rate.  Another example of new physics which
can alter the expansion rate is the late decay of a very massive 
particle which reheats the universe, but to a ``low" reheat 
temperature~\cite{lowreheat}.  If the reheat temperature is too 
low ($\la 7$~MeV) the neutrinos will fail to be fully populated, 
resulting in $\Delta$N$_{\nu} < 0$ and $S < 1$.  Finally, in some 
higher dimensional extensions of the standard model of particle 
physics additional terms appear in the 3+1 dimensional Friedman 
equation whose behavior mimics that of ``radiation", resulting 
in an effective $\Delta$N$_{\nu}$ which could be either positive 
or negative~\cite{rs}.  

Thus, a nonstandard expansion rate ($S \neq 1$) is a well-motivated, 
one parameter modification of SBBN which has the potential to resolve 
{\it some} of its challenges.  A slower expansion would leave more time
for neutrons to become protons and a lower neutron abundance at BBN
would result in a smaller \Yp (good!).  Since \4he is the most sensitive
chronometer, the effect on its abundance is most significant.  However,
a modified expansion rate would also affect the predicted abundances
of the other light nuclides as well.  A slowdown will result in more
destruction of D and \3he and, for $\eta_{10}~\ga 4$, in production 
of more $^{7}$Be (which becomes \7li via electron capture).  In 
Figure~\ref{svseta} are shown the quite accurate approximations to 
the isoabundance curves for D and \Yp in the $S - \eta_{10}$ plane, 
from the recent work of Kneller and myself~\cite{ks}.  Also shown 
in Fig.~\ref{svseta} is the location in this plane corresponding to 
the adopted D and \4he abundances (including their uncertainties).  
Not surprisingly, it is possible to adjust these two parameters 
($S$ and $\eta$) to fit the relic abundances of these two nuclides.  
Note, however, that the best fit corresponds to a {\it slower} than 
standard expansion rate ($S \approx 0.94\pm0.03$; $\Delta$N$_{\nu} 
\approx -0.70\pm0.35$).  While this combination of parameters is 
consistent with WMAP (see, \eg Barger et al.~\cite{barger} and 
further references therein), it does not resolve the conflict with 
\7li.  Although a slowdown in the expansion rate does have the effect 
of increasing \7li, this is compensated by the somewhat lower baryon 
density, which has the opposite effect.  The result is that for these 
choices of $S$ and $\eta$, which do resolve the conflicts between D 
and \4he (and WMAP and \4he), the predicted primordial abundance of 
\7li is still [Li]$_{\rm P} \approx 2.62\pm0.10$. 

\section{Summary and Conclusions}

Four light nuclides (D, \3he, \4he, \7li) are predicted to
emerge from the early evolution of the universe in abundances
large enough to be observed at present.  In SBBN there is only 
one parameter, the baryon density parameter $\eta$, which determines
the relic abundances of these nuclides.  The current observational
data identifies a range in this parameter of about a factor 
of 2-3 for which the SBBN predictions are in agreement with 
the primordial abundances inferred from current data.  This
range of $\eta$ is also consistent with independent, non-BBN
estimates~\cite{sperg}.  Tests of the standard model of cosmology 
at 20 minutes (BBN) and nearly 400 thousand years later (WMAP)
agree.  While this is a great triumph for the standard models 
of cosmology and of particle physics, the agreement is not 
perfect and, if the uncertainty estimates are taken seriously, 
there are some challenges to this standard model.  In this talk, 
to an audience of astronomers, I have emphasized the observational 
uncertainties in the hope of helping to stimulate further 
observational (and theoretical) work.  Will more and better 
data resolve these apparent conflicts?  Or, will we be pointed 
to new physics beyond the current standard models?  Only time 
will tell.

\newpage

\noindent{\bf Acknowledgments}

It is with great pleasure that I express my thanks to the organizers
of this very interactive and stimulating meeting, especially Luca
Pasquini and Sofia Randich for their tireless efforts to facilitate
my participation and to make it so enjoyable.  The research described
here has been supported at The Ohio State University by the US Department
of Energy (DE-FG02-91ER40690).

%

\end{document}